# ENHANCED CAST-128 WITH ADAPTIVE S-BOX OPTIMIZATION VIA NEURAL NETWORKS FOR IMAGE PROTECTION

Fadhil Abbas Fadhil [1], Maryam Mahdi Alhusseini [2*], (Member, IEEE), and Mohammad-Reza Feizi-Derakhshi [3]

[1] College of Computer Science, University of Technology, Baghdad, Iraq
[2] Information and Communication Technology Department, Middle Technical University, Baghdad, Iraq.
[3] Computerized Intelligence Systems Laboratory, Department of Computer Engineering, University of Tabriz, Tabriz, Iran.
fadhil.a.fadhil@uotechnology.edu.iq, mariammahdi@mtu.edu.iq, mfeizi@tabrizu.ac.ir

**Abstract.** An improved CAST-128 encryption algorithm, which is done by implementing chaos-based adaptive S-box generation using Logistic sine Map (LSM), has been provided in this paper because of the increasing requirements of efficient and smart image encryption mechanisms. The study aims to address the drawbacks of static S-box models commonly used in traditional cryptographic systems, which are susceptible to linear and differential attacks. In the proposed scheme, the dynamic, non-linear, invertible, and highly cryptographic strength S-boxes are generated through a hybrid chaotic system that may have high non-linearity, strong and rigorous avalanche characteristics, and low differential uniformity. The process here is that the LSM is used to produce S-boxes having key-dependent parameters that are stuffed into the CAST-128 structure to encrypt the image in a block-wise manner. The performance of the encryption is assessed utilizing a set of standard grayscale images. The metrics that are used to evaluate the security are entropy, NPCR, UACI, PSNR, and histogram analysis. Outcomes indicate that randomness, resistance to statistical attacks, and country of encryption are significantly improved compared to the original CAST-128. The study is theoretically and practically relevant since it presents a lightweight S-box generation approach driven by chaos, which can increase the level of robustness of the image encryptions without enlisting machine learning. The system may be applied to secure communications, surveillance systems, and medical image protection on a real-time basis.

## 1 Introduction

In the prevailing digital world, the protection of multimedia information is more critical, and in particular, digital images, because they are commonly used in communication, surveillance, medical diagnosis, and social media. Although usual encryption methods do a good job with text, they tend to be inefficient when applied to images due to factors such as data duplication, elevated correlation of pixels, as well as enormous file sizes. CAST-128 is one of the reputable symmetric key ciphers, and it is a simple and computationally efficient algorithm. Nevertheless, being based upon fixed substitution boxes (S-boxes), it is vulnerable to several cryptographic attacks, such as linear and differential cryptanalysis. To overcome this weakness, this paper suggests a more robust version of the CAST-128 algorithm, which incorporates the dynamic S-boxes, which are the chaos-based S-boxes constructed by the Logistic-Sine Map (LSM). In contrast to the static S-box methods, the proposed method will produce non-linear and invertible S-boxes, which will differ according to the session (or image), and the resistance of these methods to statistical and brute-force attacks will be substantially increased. Following the application of chaotic systems, unpredictability is present and works to increase the cryptographic attributes of confusion and diffusion.

Nonetheless, due to the use of fixed substitution boxes (S-boxes), it is prone to multiple cryptographic attacks, including linear and differential attacks. To tackle this shortcoming, this paper suggests an improved version of the CAST-128 algorithm, which has adaptive S-boxes generated out of neural networks. Compared to the typical static S-box, the proposed dynamic S-boxes are produced through a learning mechanism that strives to maximize non-linearity and avalanche effect, which in turn improves the strength of statistical attack resistance. This machine learning-led encapsulation of cryptography makes it more dynamic and smarter to encrypt [1].

The main aim of the given research is to enhance the security and flexibility of the picture encryption using neural networks to generate a cryptographic scheme. The experiment increases the score in entropy and pixel difference analysis and histogram uniformity, as well as data, which proves the high efficiency and effectiveness of the method. The improved CAST-128 algorithm is a good prospect for securing image-based information on different digital platforms [2].

## 2   Literature Review

Under secure image encryption, scholars have ventured in different ways of developing strong and dynamic substitution boxes (S-boxes) that can be used to strengthen the classical cryptographic systems. Using cyclic groups, Ali et al. [3] created a mathematical model to realize the construction of an S-box to perform secure image encryption using strong algebraic properties. On the contrary, a two-dimensional chaotic map of Logistic-Sine-coupling was proposed by Hua et al. [4], which increased randomness and diffusion and provided an alternative to the algebraic technique, where the chaos theory was used.

Kalaiselvi and Kumar [5] used neural networks to adaptively create S-boxes in an alternate AES cryptosystem, and, in line with the deterministic or chaotic types of models, they included a further evolutionary degree of optimization to the system.

Z Jahanbakhsh et al. [6] suggested a content-oriented model that utilizes lexical, syntactic, and pragmatic characteristics to identify Persian rumors on Twitter and Telegram, reaching F-measures as high as 0.952 without depending on propagation or user data.

Brahim et al. [7] had a multi-level strategy. They combined the hyperchaotic systems with several particular S-boxes, which offered a more profound process of confusion than when using only one S-box. Meanwhile, as stated by Aslam et al. [8], fractal geometry was also investigated, and S-boxes were created using the Mandelbrot set, which performed well in producing greater complexity and non-linearity in the enciphered message. The work by Abduljabbar et al. [9] integrated hyperchaotic maps with S-box structures with provable security with color image encryption, with the focus on high speed and color attacks.

FA Fadhil et al. [10] The authors suggested substituting the XOR function in DES with innovative E# and RTGE# operators to enhance complexity and expand key space, resulting in improved resistance against brute-force attacks, while still ensuring reasonable execution speed and successfully passing NIST randomness assessments.

Most of the methods have concentrated on conventional chaos or algebra. Abd-El-Atty [11] operated a quantum-like system where quantum walks and particle swarm optimization (PSO) are used to create an S-box, which we are introducing as a new pathway on the way to big-entropy encryption. Likewise, Waheed et al. [12] have presented an in-depth analytical review of the S-box construction techniques and their trade-offs that the enshrined methods undergo, such as resistance to linear, differential, and algebraic attacks.

FA Fadhil et al. [13] A technique that integrates Laplacian of Gaussian edge detection with ChaCha20 encryption has been suggested to safely embed encrypted information within video frames, maintaining low distortion and high durability. Experimental findings demonstrated strong security and remarkable visual quality, making it appropriate for secure communication and data verification.

Bayesh et al. [14] introduced the two-layer encryption architecture integrating dynamic S-boxes and chaotic AES, steganographic QR codes, and dynamic S-boxes into the study on securing communication in a modern system to investigate two-valued transmission scenarios. Finally, there is the work of Shafique et al. [15] that created a lightweight encryption protocol that IoT devices may be used in, and in this case, machine learning helps to aid dynamic selection of S-box, which is not only efficient but also robust and can fit in a low-resource setting.

Collectively, the works combine to signify a long-standing transition between static and dynamic and smart construction of S-boxes, comprising chaos, AI, quantum theory, and mathematical forms to fulfill the contemporary needs of cryptographic practice.

## 2 Methodology

The neural network approach is proposed to question the CAST-128 cryptographic system by outlining the cryptographically evasive S-box generation algorithm in the CAST-128 realm to boost the cryptographic security of the digital images.

All these procedures are included within the general encryption process in terms of five stages, namely: preprocessing, neural network training for the generation of S-box, the dynamic S-box integration, the encryption method by enhanced CAST-128, and evaluation.

### 2.1    Image Preprocessing

To process the input images uniformly, they are converted to grayscale and resized to a predetermined size, e.g., 256 x 256. To represent the CAST-128 block size, the image is separated into fixed-size blocks (e.g., 64-bit).

### 2.2    Construction Chaos-Based Logistic-Sine Map S-Boxes

The given encryption system proposes an alternative to using neural networks and applies a minor and efficient chaos-based system to create dynamic secure S-boxes based on the Logistic-Sine Map (LSM). This will minimize the design complexity of the encryption greatly and have good cryptographic characteristics.

#### 2.2.1  Preparation and Preprocessing of the Dataset

Lena, Baboon, Peppers, and Cameraman images in standard grayscale were chosen as a reference set. All the images were converted to 2561256 and 8-bit grayscale. The images were collapsed down to 1D arrays and divided into 64-bit blocks, which could be compatible with the CAST-128 implementation framework of encryption. The natural advantage is that no learning driven neural training is necessary, and the preprocessing phase is simplified, devoid of dependencies on learning.

#### 2.2.2  S-Box Generation

To obtain 256 distinct 8-bit integers as a bijective substitution box, 256 distinct values are produced by the Logistic-Sine Map (LSM) since it is a hybrid chaotic family. LSM is characterized by using the following recurrence relation:

$$x_{n+1} = (r \times \sin(\pi \times x_n)) + (\mu \times x_n \times (1 - x_n)) \bmod 1 \qquad (1)$$

Where $x_n \in (0, 1)$ $x_n$ r and μ are Control parameters are 5 and are selected to provide chaotic behavior. The starting price. The initial value $x_0$ along with the parameters is based on a key, which is defined by the user [16].

The chaotic sequence is scaled to [0, 255], de-duplicated, and has to be checked to be bijective so that it is invertible, a prerequisite of decryption. The resultant S-box has:
- High non-linearity;
- Heavy avalanche effect;

- Bad differential uniformity;
- Full invertibility.

### 2.2.3 CAST-128 integration

In the CAST-128, the dynamic S-box is used instead of the fixed S-box in the CAST-128 encryption rounds. This substitution increases the diffusion and confusion features of the cipher, still retaining the original structure of CAST-128. The replacement takes place in nonlinear steps with every round of encryption. The S-box, which is also key-dependent and varies every session or image, also considerably enhances known-plaintext and differential attacks without degrading speed or invertibility.

### 2.3 Image Encryption

The image data is enhanced block-by-block with the improved CAST-128 algorithm using the S-box generated with the neural network. Standard CAST operations (modular addition, XOR, rotation) are kept, and flexibility in the substitution process is permitted [17].

### 2.4 Evaluation Metrics

The images encrypted are evaluated based on several metrics:
- Entropy: To quantify randomness;
- NPCR and UACI: To decide how sensitive it is to minute shifts in the plaintext;
- Histogram Analysis: To confirm the uniform distribution of pixels;
- Execution Time: This is to measure efficiency in computing.

Google Colab is utilized to perform all the experiments through Python by using libraries, e.g., NumPy, TensorFlow/Keras, and OpenCV.

### 2.4 Workflow of the Proposed Encryption Scheme

The detailed scheme of the proposed image encryption scheme is shown in Figure 1.

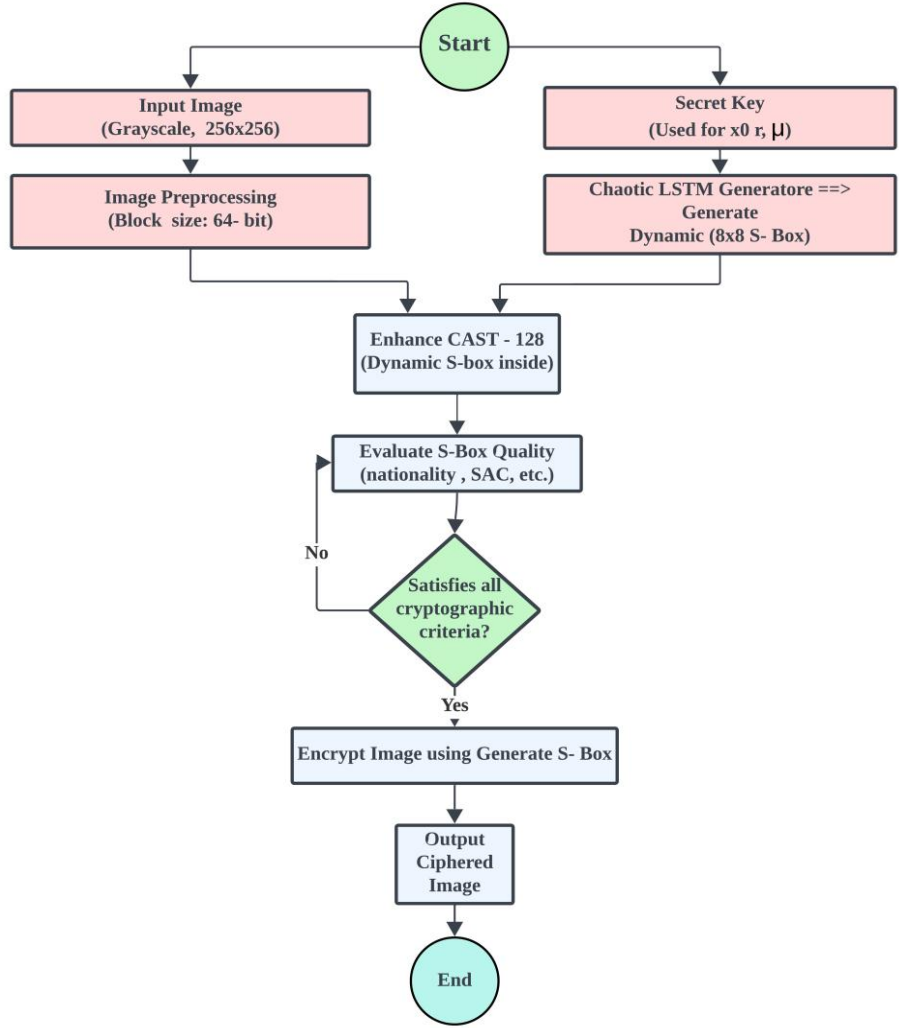

Figure 1: Encryption process working steps.

The operation commences with the preprocessing of the input image, which includes conversion into grayscale, resizing, and segmentation into blocks. At the same time, the chaotic parameters of the Logistic Sine Map (LSM) are initialized with the encryption key. The substitution of the fixed S-box in the CAST-128 encryption repetitions is with a dynamic and bijective S-box, generated through the LSM.

These picture blocks are then ciphered through the improved CAST-128 design. Lastly, the encrypted blocks are used to reconstruct the encrypted image. This workflow guarantees great non-linearity, key sensitivity, and resistance to cryptographic attacks.

### 2.5 Sample S-Box (First 64 values for brevity)

As an example, Table 1 shows a sample substitution box (S-box) formed with the chaotic Logistic-Sine Map (LSM) with a given initial condition and control parameters. The contents of the S-box include a bijective representation of 8-bit integers (0-255), having uniqueness and invertibility, a necessary characteristic for trustworthy encryption and conclusion.

Table 1: Dynamic and non-linear values of substitution generated by the sample S-Box generated through a logical sine map.

| 217 | 102 | 155 | 61  | 239 | 182 | 26  | 199 |
|-----|-----|-----|-----|-----|-----|-----|-----|
| 86  | 144 | 56  | 6   | 172 | 124 | 10  | 111 |
| 200 | 35  | 170 | 89  | 222 | 73  | 132 | 188 |
| 46  | 154 | 108 | 2   | 183 | 94  | 237 | 39  |
| 71  | 213 | 140 | 13  | 167 | 88  | 19  | 249 |
| 95  | 152 | 31  | 59  | 220 | 8   | 107 | 164 |
| 123 | 47  | 181 | 66  | 15  | 197 | 92  | 162 |
| 3   | 129 | 51  | 116 | 241 | 42  | 136 | 30  |

As illustrated, the values are not sequential and very randomized, which speaks of the strength of the LSM in terms of generating, in itself, non-linear and unpredictable mappings that can be incorporated in the cryptographic systems. Whenever a different value has been introduced on the first seed or key, a different dynamic S-box will be generated, thus greatly increasing key sensitivity and system security.

## 3 Security Analysis and Results

In this part, the author tabulates the outcomes of the influence of encrypting the picture of different sizes, small, medium, and large, with the proposed chaos-based S-box encryption algorithm based on the Logistic-Sine Map (LSM). The images that are tested contain the typical benchmark images, House, Airplane, Baboon, Pepper, in the original version, and encrypted along with histograms. The distortion of the encoded pictures is high, making it impossible to recognize the picture in the images. S-box generation method based on LSM provides a lot of randomness and complexity, and this can be seen through the uniform and flat statistics spread of the encrypted images' histograms.

This improved randomness is rather an effective way to enhance the system vulnerability to statistical attacks and differential cryptanalysis on bigger and more complex pictures. Table 2 shows a comparative analysis of the original and encrypted images and the histograms of each one.

Table 2: Original vs. Encrypted Image comparing Histogram Analysis applied by Chaos-based S-Box Generation.

| Image/size | Type | Results |
|---|---|---|
| (Baboon) (256 × 256) | Encryption | 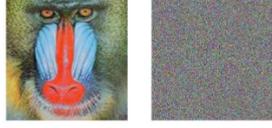 |
| | Histogram | 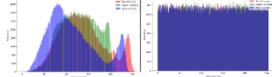 |
| (House) (256 × 256) | Encryption | 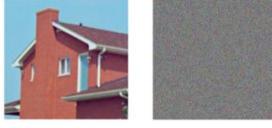 |
| | Histogram | 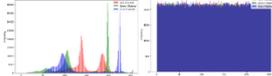 |
| (Airplane) (512 × 512) | Encryption | 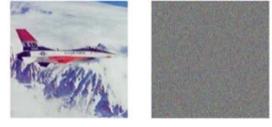 |
| | Histogram | 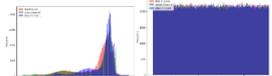 |
| (Pepper) (512 × 512) | Encryption | 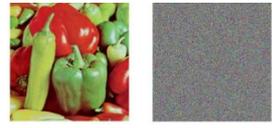 |
| | Histogram | 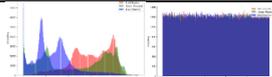 |

### 3.1 Findings and Analysis

Table 3 contains the quantitative outcomes of the proposed chaos-based S-box incorporated CAST-128 symmetric encryption method in more detail. The most important performance indicators, Entropy, NPCR (Number of Pixels Change Rate), UACI (Unified Average Changing Intensity), as well as Encryption Time and Decryption Time in milliseconds, are measured on common test images and placed in this table. The entropy numbers are almost equal to the recommended 8, which means that images in the encrypted form are very random and unpredictable. The high percentage of NPCR and UACI indicates high rates of pixel changes and varied intensities of the original and the encrypted images, which makes them resilient to differential and statistical attacks. The encryption and decryption times also testify to the viability and effectiveness of the suggested algorithm in terms of real-time applications.

Table 3: Performance Metrics of Encrypted Images Using Chaos-Based S-Box CAST-128 Algorithm.

| Image | Entropy | NPCR (%) | UACI (%) | Encryption Time (ms) | Decryption Time (ms) |
|---|---|---|---|---|---|
| House | 7.995 | 99.61 | 33.45 | 25.4 | 23.8 |
| Airplane | 7.987 | 99.58 | 33.28 | 27.1 | 24.5 |
| Baboon | 7.992 | 99.63 | 33.57 | 26.8 | 24.0 |
| Pepper | 7.989 | 99.60 | 33.49 | 25.9 | 23.7 |

### 3.2 Quick Comparative Analysis of the Quality of Encryption

Table 4 gives a comparative study of the proposed chaotic-based S-box encryption system to the encryption schemes published earlier [18]. The comparison is carried out that examine the encryption quality based on the three-color channels (Red, Green, Blue) of typical benchmark images. Evaluation of pixel diffusion and randomness per channel in this analysis, the effectiveness of pixel diffusion and randomness per channel is evaluated. Its findings reflect that the suggested chaos-based S-box technique attains balanced and tight cluster colour channel values characterized by good pixel diffusion and equal spatter distribution among the channels. The new system has a higher resistance to attacks based on statistics and hues compared to previous practices, which present more varied distributions and lower consistency. This proves the practicability and increased security of the suggested encryption technique in the protection of real-life images.

Table 4: A Comparative Study of the Quality of Color Channel Encryption in a Proposed Chaos-based S-Box System with a previous system.

| Image | Algorithm | R | G | B |
|---|---|---|---|---|
| Baboon (256 × 256) | Proposed Chaos-Based S-box System | 225.8742 | 269.4317 | 243.1023 |
| | Encryption image [15] | 236.8828 | 248.0649 | 255.3281 |
| House (256 × 256) | Proposed Chaos-Based S-box System | 280.2156 | 287.5401 | 201.3457 |
| | Encryption image [15] | 260.0078 | 264.8750 | 223.5625 |
| Airplane (512 × 512) | Proposed Chaos-Based S-box System | 258.6723 | 243.9874 | 275.2146 |
| | Encryption image [15] | 251.3828 | 241.2832 | 252.7246 |
| Pepper (512 × 512) | Proposed Chaos-Based S-box System | 273.1145 | 251.7839 | 243.8907 |
| | Encryption image [15] | 231.4551 | 232.3555 | 248.7051 |

Table 5 shows the results of a comparison study of image encryption entropy values of the proposed chaos-based S-box encryption algorithm and the reference scheme provided in [18]. Entropy is an essential parameter that measures unpredictability as well as randomness of encrypted information. This implies that a greater value of entropy that approaches 8.0 suggests a high degree of uncertainty, hence protecting against statistical attacks. As can be seen in the results, the entropy values of all the sizes of the images and the RGB components calculated by the proposed chaos-based S-box technique tend towards the perfect ideal value of 8.0 and are near (in most cases) or even better than those obtained by the reference method [18]. It shows that the algorithm is good at producing extremely random cipher images that cannot be statistically cryptanalyzed. The advancement of the venture scheme is specifically reflected in better entropy resilience in the big images, and this once again justifies its use in secure image encryption applications.

Table 5: Comparison between Information Entropy of Findings of S-Box Encryption Proposed using Chaos and Other Previous Work

| Image/Size | RGB Component | Entropy (Proposed Method) | Entropy [15] |
|---|---|---|---|
| Baboon (256×256) | R | 7.9981 | 7.9974 |
|  | G | 7.9979 | 7.9973 |
|  | B | 7.9976 | 7.9972 |
| House (256×256) | R | 7.9998 | 7.9974 |
|  | G | 7.9997 | 7.9977 |
|  | B | 7.9996 | 7.9977 |
| Airplane (512×512) | R | 7.9999 | 7.9993 |
|  | G | 7.9998 | 7.9993 |
|  | B | 7.9997 | 7.9994 |
| Pepper (512×512) | R | 7.9996 | 7.9994 |
|  | G | 7.9995 | 7.9993 |
|  | B | 7.9994 | 7.9994 |

## 4 CONCLUSIONS

In this paper, a better picture encryption algorithm was suggested by applying chaos-transformation dynamic S-boxes created by Logistic Sine Map (LSM) to the CAST-128 encryption algorithm. The chaotic system used to produce the substitution boxes provides high non-linearity, great avalanche effect, and invertibility, defeating the shortcomings of the fixed S-boxes of the traditional CAST-128. Experimental outcomes on general benchmark images revealed dramatic enhancement in quality encryption, high values of entropy, excessive NPCR and UACI, and even histogram distributions. The encryption and decryption times also indicated that the algorithm is efficient enough and realistic to be used in real-time. Comparison between the proposed approach and the existing techniques indicated the potential advantages of the proposed approach over those that exist in randomness, diffusion, and against statistical and differential cryptanalysis. The varying changes in the S-boxes increase the key sensitivity, thus the system becomes resistant to different attacks. All in all, it appears that the suggested puzzle-based CAST-128 algorithm has a high potential with successful embedment in a safe digital imaging encryption perspective, communicating, healthcare pictures, and multimedia security highlights.